\documentclass[useAMS,usenatbib]{mn2e}
\usepackage{graphicx}

\title[${[Fe/H]}$ relations for c-type RR Lyrae Variables]
{${\rm\bf[Fe/H]}$ relations for c-type RR Lyrae Variables based upon Fourier Coefficients}
\author[S. M. Morgan, J. N. Wahl and R. M. Wieckhorst]{Siobahn M. Morgan,\thanks{E-mail:
siobahn.morgan@uni.edu} Jennifer N. Wahl and Rachel M. Wieckhorst\\
Department of Earth Science, University of Northern Iowa, Cedar Falls, Iowa, 50614-0335, USA}

\begin{document}
\date{Accepted 0000 December 00. Received 0000 December 00; in original form 0000 October 00}

\pagerange{\pageref{firstpage}--\pageref{lastpage}} \pubyear{0000}

\maketitle

\label{firstpage}

\begin{abstract}
[Fe/H] - $\phi_{31}$ - $P$ relations are found for c-type 
RR Lyrae stars in globular clusters. The relations are 
analogous to that found by \citet{jk96} for 
field ab-type RR Lyrae stars, where a longer period 
correlates with lower metallicity values for similar values 
of the Fourier coefficient $\phi_{31}$. The relations
obtained here are used to determine the metallicity of 
field c-type RR Lyrae stars, those within $\omega$ Cen, 
the LMC and toward the galactic bulge.  The results are found to compare favorably to 
metallicity values obtained elsewhere. 
\end{abstract}

\section{Introduction}

For nearly one hundred years, Fourier functions have been used to determine
the pulsation period, $P$, of regular variables.  
\citet{sl81} were able to show that other useful information 
apart from the period could be derived from Fourier functions.  
Their format of the relation is commonly used: 
\begin{equation}
V(t) = A_0 + \sum_{i=1}^{n} A_i \cos (i \omega t + \phi_i)
\label{eqn1}
\end{equation}
where the terms $A_i$ and $\phi_i$ are the Fourier coefficients of the fit of degree $n$, and 
$\omega=2\pi/P$.  Generally the values of $A_i$ and $\phi_i$ were ignored when
this function was used to determine the value of $P$.
\citet{sl81} showed that these terms when combined in the following manner --
$R_{ij}=A_i/A_j$ and $\phi_{ij}=j\phi_i-i\phi_j$, 
could provide more information about pulsating variables than at first thought.
Currently, the coefficients derived from the light curves of 
Cepheids and RR Lyrae have been used to provide information 
about pulsation modes and resonance effects, as well as physical characteristics
of the stars such as mass, luminosity, and metallicity.

Jurcsik \& Kov\'{a}cs (1996, hereafter JK96) surveyed field ab-type RR Lyrae (RRab) and 
derived a relationship between $P$, $\phi_{31}$ and [Fe/H] 
for these variables which has been used in a variety of applications,
in particular to determine the metallicity of 
stars found in large scale surveys (Morgan, Simet \& Bargenquast 1998, hereafter MSB98), 
in globular clusters (Cacciari, Corwin \& Carney 2005) 
and in other galaxies \citep{detal05}.
\citet{s04} describes this relationship as an aspect of the Oosterhoff--Arp--Preston
period-metallicity effect, in which the apparent shift of the value of $P$ or
$\phi_{31}$ is dependent upon the influence metallicity has on  
horizontal branch morphology.  \citet{kw01} expanded upon the work of JK96 by investigating
possible relationships between the physical parameters of RRab stars
and other Fourier coefficients.  They found several trends in the value of 
photometric colours and magnitudes with the coefficients.

The utilization of the Fourier coefficients of the c-type RR Lyrae variables (RRc) has not been as 
extensive as that of the RRab stars.
\citet{s89} used hydrodynamic models of RRc stars to show relationships exist between the
values of helium abundance ($Y$), luminosity, mass, $P$ and $\phi_{31}$.  In particular, he 
found that $\phi_{31}$ is directly related to the luminosity-mass ratio, $L/M^{1.81}$,
for these stars.  \citet{s90} showed that the distribution of $\phi_{31}$ with period would vary
according to Oosterhoff group, with Oosterhoff I cluster stars having larger values of  
$\phi_{31}$ for a given value of $P$ than stars from Oosterhoff II clusters.
Clement, Jankulak \& Simon (1992) found strong evidence for the trend of 
$\phi_{31}$ increasing with period in their observations
of several globular clusters, and also noted the trend for lower metallicity clusters
to have their $\phi_{31}$ values shifted to longer periods than the higher metallicity
clusters on a $\phi_{31}$-$P$ diagram.  This result was also observed in 
the RRc variables in NGC 4590 (M68) by Clement, Ferance \& Simon (1993).  The general nature of this
aspect of the variation of $\phi_{31}$ with metallicity has previously been used 
to estimate the metallicity of RRc stars in the OGLE surveys
of the galactic bulge (MSB98) and toward 47 Tuc \citep{md00},
however these results were at best only general approximations of [Fe/H].

The lack of a metallicity relationship for the RRc stars analogous to that derived by
JK96 is likely due to the scarcity of accurate metallicity and light curve data 
for field RRc variables.  
Generally metallicities for field RRc stars are estimated using the $\Delta S$ method
of \citet{preston59}, 
with only a few metallicities derived from high resolution spectroscopy 
\citep{betal82,letal96,fb97,setal97}.  
In several cases, $\Delta S$ values obtained for a single RRc star can
vary significantly, making the accuracy of the metallicity for such stars suspect.

Fortunately, there is an abundance of data for globular cluster RRc variables, including values for
cluster metallicity and a significant number of Fourier coefficients.  We will use these
data to derive relationships for RRc stars relating the Fourier coefficient $\phi_{31}$, pulsation period
and [Fe/H].  The resulting relations will also be applied to several test
cases, including field RRc stars, as well as variables in $\omega$ Cen, the
Large Magellanic Cloud and toward the galactic bulge.

\section{Method}

The Fourier coefficients for globular cluster RRc variables were obtained from the Fourier coefficient
website \citep{m03} as well as several recent publications.  The 
clusters used here are listed in Table~\ref{tbl-1}.  In order to avoid possible errors 
with conversion from one photometric
system to another, only Fourier coefficients derived from $V$ magnitudes were considered.  
Where ever possible the light curve data for the individual stars 
were examined and Fourier fits that were based upon sparsely sampled light curves
or with poorly defined maxima/minima were excluded.  Fourier coefficients with
large uncertainties were also excluded.  The sources for the Fourier coefficients, 
and the number of stars from each cluster ultimately used in this study are included 
in Table~\ref{tbl-1} as well.
There are several sources for metallicity that could be used for globular clusters.
Two metallicity scales that are frequently cited in the literature
are those of Zinn \& West (1984, hereafter ZW84) and Carretta \& Gratton (1997, hereafter CG97).
Both of these metallicity scales will be used here.  For some clusters, 
the updated values from \citet{zinn85} are used in place of the ZW84 values
where available.  When values were not available for clusters based on the system of CG97, the values 
found by ZW84 were converted to the CG97 scale using the relation of \citet{cetal01}.  The
metallicity values and their corresponding uncertainties are also given in Table~\ref{tbl-1}.

\begin{table}
\caption{Globular Clusters containing RRc Stars with reliable Fourier Coefficients.}
\label{tbl-1}
\begin{tabular}{lccccc}
\hline
Cluster & Symbol & Source & N & $\rm[Fe/H]_{ZW}$ & $\rm[Fe/H]_{CG}$ \\
\hline
NGC 6171 & $\times$ & 1 & 7 & $-0.99\pm0.06$ & $-0.97\pm0.04^\ast$ \\
NGC 6362 & $\times$ & 2 & 13 & $-1.08\pm0.09$ & $-0.96\pm0.01$ \\
NGC 1851 & $\bigcirc$ & 3 & 4 & $-1.33\pm0.09$ & $-1.18\pm0.05^\ast$ \\
NGC 5904 & $\bigcirc$ & 4 & 14 & $-1.40\pm0.06$ & $-1.11\pm0.11$ \\
NGC 6934 & $\bigtriangleup$ & 5 & 6 & $-1.54\pm0.09$ & $-1.32\pm0.07^\ast$ \\
NGC 7089 & $\bigtriangleup$ & 6 & 3 & $-1.62\pm0.07$ & $-1.38\pm0.06^\ast$ \\
NGC 5272 & $\bigtriangleup$ & 7 & 17 & $-1.66\pm0.06$ & $-1.34\pm0.06$ \\
NGC 6333 & $+$ & 8 & 5 & $-1.78\pm0.15$ & $-1.52\pm0.16^\ast$ \\
NGC 4147 & $+$ & 9 & 8 & $-1.80\pm0.26$ & $-1.55\pm0.28^\ast$ \\
NGC 6809 & $+$ & 10 & 5 & $-1.82\pm0.15$ & $-1.57\pm0.17^\ast$ \\
NGC 4590 & $\bigtriangledown$ & 11 & 14 & $-2.09\pm0.11$ & $-1.99\pm0.10$ \\
NGC 7078 & $\bigtriangledown$ & 12,13 & 10 & $-2.15\pm0.08$ & $-2.12\pm0.01$ \\
\hline
\end{tabular}

\medskip
Sources: 
1 - \citet{cs97}, 2 - \citet{oetal01},
3 - \citet{w98}, 4 - \citet{ketal00}, 
5 - \citet{ketal01}, 6 - \citet{aetal04},
7 - \citet{ccc05}, 8 - \citet{cs99},
9 - \citet{setal05}, 10 - \citet{oetal99},
11 - \citet{w94}, 12 - \citet{cs96}, 
13 - \citet{aetal06} \\
$\ast$ - calculated using \citet{cetal01}
\end{table}

\begin{figure}
  \includegraphics[width=84mm]{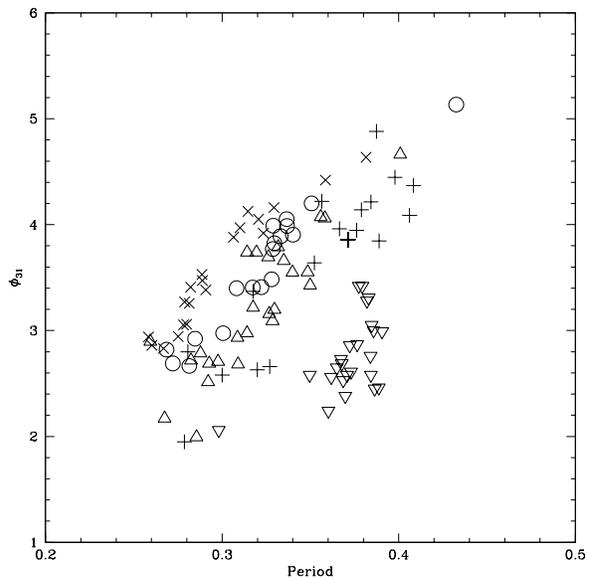}
\caption{Fourier coefficient $\phi_{31}$ plotted versus period for the
RRc stars in the globular clusters listed in Table~\ref{tbl-1}.  The various
symbols are based upon the cluster's value for [Fe/H].}
\label{fig1}
\end{figure}

Overtone pulsators such as RRc stars generally have a lower 
order fit (small value of $n$ in equation~\ref{eqn1}) used for the light curve,
which limits the number of Fourier coefficients available.  
As was outlined in the introduction and shown in \citet{cjs92} and MSB98, 
the coefficient that appears to depend strongest upon [Fe/H] is the $\phi_{31}$ term.  
This trend is also observed in RRab stars as was noted by JK96 and \citet{cs99}. 
Our study concentrated exclusively on this coefficient.  
The $\phi_{31}$ values for the $106$ stars in our sample are plotted relative to $P$ 
in Figure~\ref{fig1}, with the coding in the diagram based upon the [Fe/H] values from Table~\ref{tbl-1}.  

\begin{figure}
  \includegraphics[width=84mm]{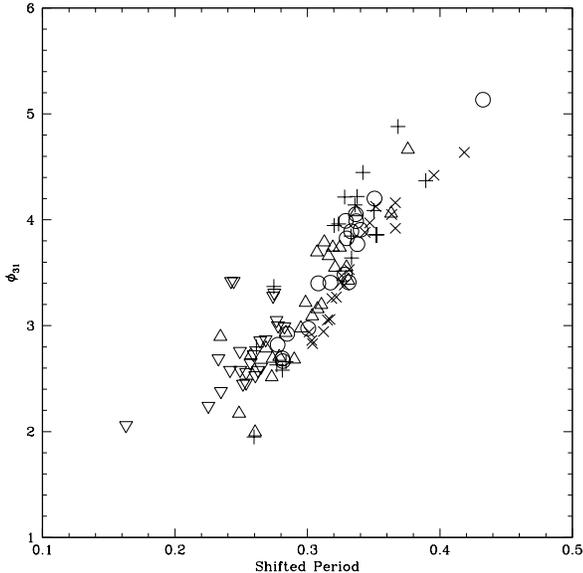}
\caption{Fourier coefficient $\phi_{31}$ plotted versus the shifted value of $P$.  
All variables within a given globular cluster are shifted by the same amount in $P$.}
\label{fig2}
\end{figure}

\begin{figure}
  \includegraphics[width=84mm]{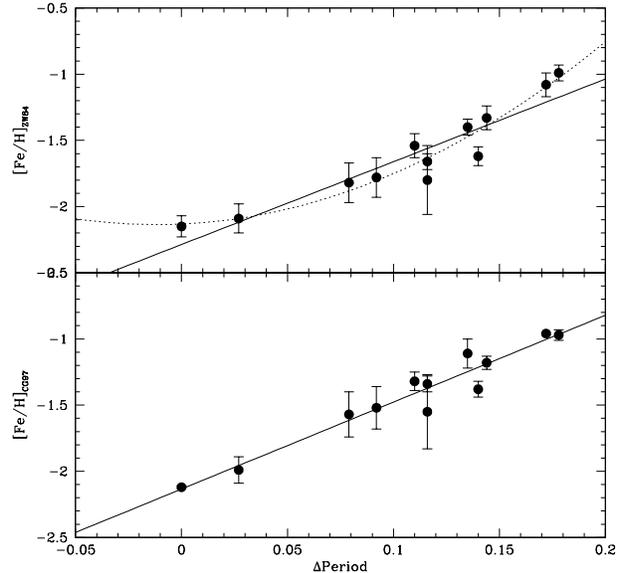}
\caption{The amount of shift in period used to produce the distribution
shown in Figure~\ref{fig2} for each cluster plotted 
versus the cluster's value of [Fe/H] for the ZW84 (top) and CG97 (bottom) systems.  
The best fit quadratic and linear relations are drawn through the ZW84 data,
while only a linear relation is shown for the CG87 system.}
\label{fig3}
\end{figure}

The relation between [Fe/H] - $P$ - $\phi_{31}$ found by JK96 for RRab stars is linear as is 
a similar relation found by \citet{s04}.  It should be noted that \citet{s04} used $\log P$ 
rather than $P$ with virtually the same degree of accuracy as that obtained with the JK96 relation.  
We found the use of a $P$ term to be slightly better than the use of a $\log P$ 
term when comparing the residuals of the derived relations to the observed values.  
The form of the relationship between [Fe/H], $P$ and $\phi_{31}$ appears to 
depend upon the metallicity scale that is used, ZW84 or CG97.  This is seen 
when the $P$ values for the RRc stars in each cluster are increased by the same amount to place
the $\phi_{31}$ values along a common distribution, as is illustrated in Figure~\ref{fig2}. 
The dependence of the $\phi_{31}$ values on $P$ is apparent, while the amount that was added to 
the period of each cluster's RRc stars indicates the dependence of $P$ on the cluster's metallicity.
The amounts added to the periods of the RRc in each cluster in order to produce the distribution in 
Figure~\ref{fig2} are shown in Figure~\ref{fig3}.
For the CG97 scale, a linear relation appears to 
exist between the shift in $P$ and [Fe/H], while
for the ZW84 scale a non-linear relation appears to be appropriate.
In the case of the ZW84 data, the standard deviation of the points from a linear fit is
$\sigma = 0.127$, while a quadratic fit has a standard deviation of $\sigma = 0.092$.
Applying a quadratic fit to the CG97 data results in a function that is nearly
identical to the linear one, indicating that no improvement to the fit
is made with a higher order function.  Even though the CG97 scale does not 
appear to require a high-order non-linear relationship
between [Fe/H] and $P$, those relations were nonetheless examined.

In order to determine the best relationship between the three variables, all
possible permutations were made between them with functions of the form
\begin{equation}
{\rm [Fe/H]}  =  a~ P^2 + b~ P +c~ \phi_{31}^2 +d~ \phi_{31} +e~ P \phi_{31} + f,
\label{feh}
\end{equation}
tested for quality of fit to the data.  All possible combinations
of zero and non-zero values for the coefficients $a$ -- $f$ were examined using
both the ZW84 and CG97 scales.  Only relations which included at least one
$P$ and $\phi_{31}$ term were considered.  A summary of the resulting fits to
the various formulae are shown in Table~\ref{tbl-2}.  This table shows 
the range of values for the sample standard deviation of the best fitting formulae to  
the $106$ data points.
The best solution for the ZW84 scale is one with all six terms used in equation~\ref{feh}, while
the quality of the best solutions for the CG97 scale does not vary significantly when fewer
terms are used in equation~\ref{feh}.   This result is expected given the linear
relation of the [Fe/H] values on $P$ as is shown in Figure~\ref{fig3} for the CG97 system.

\begin{table}
\caption{Sample Standard Deviations ($\sigma$) for solutions to equation~\ref{feh}.}
\label{tbl-2}
\begin{tabular}{lcc}
\hline
Metallicity & Terms & Range of $\sigma$ \\
\hline
ZW84 & 6 & 0.145  \\
  & 5 & 0.152 -- 0.161  \\
  & 4 & 0.160 -- 0.170  \\
  & 3 & 0.162 -- 0.174  \\
\hline
CG97 & 6 & 0.142 \\
  & 5 & 0.142 -- 0.144  \\
  & 4 & 0.142 -- 0.154  \\
  & 3 & 0.143 -- 0.169  \\
\hline
\end{tabular}

\end{table}

Two criteria were used for determining which formulae would be the most useful for
calculating RRc metallicities.  First was the quality of the fit to the data, which is summarized in Table~\ref{tbl-2}.
The second criterion was the simplicity of the formula.  This criterion was only
relevant for the CG97 solutions, where the quality of the formula varied insignificantly
when the number of terms used in the solution were changed.
The best fit formula for equation~\ref{feh} to the clusters
in Table~\ref{tbl-1} using the ZW84 scale is
\begin{eqnarray}
{\rm [Fe/H]_{ZW}} & = & 52.466 P^2 -30.075 P + 0.131 \phi_{31}^2 \nonumber \\
 & &  + 0.982 \phi_{31} - 4.198 \phi_{31} P  + 2.424
\label{zwfit}
\end{eqnarray}
which has a sample standard deviation of $0.145$ dex.  The best formula based upon the CG97 scale is
\begin{equation}
{\rm [Fe/H]_{CG}}  =  0.0348 \phi_{31}^2 +0.196 \phi_{31} - 8.507 P + 0.367
\label{cgfit}
\end{equation}
which has a sample standard deviation of $0.142$ dex.  
$84\%$ and $87\%$ of the [Fe/H] values based upon the above formulae  
are within $0.2$ dex of the ZW84 and CG97 cluster metallicity values 
respectively, while $58\%$ and $62\%$ of the [Fe/H] values are within $0.1$ dex.  
The average difference between the [Fe/H] values
from ZW84 and CG97 and those based upon the above relations is approximately $0.14$ dex, indicating
that these relations provide values for the metallicity that are within the range of
uncertainty found using a variety of other methods.

\begin{figure}
  \includegraphics[width=84mm]{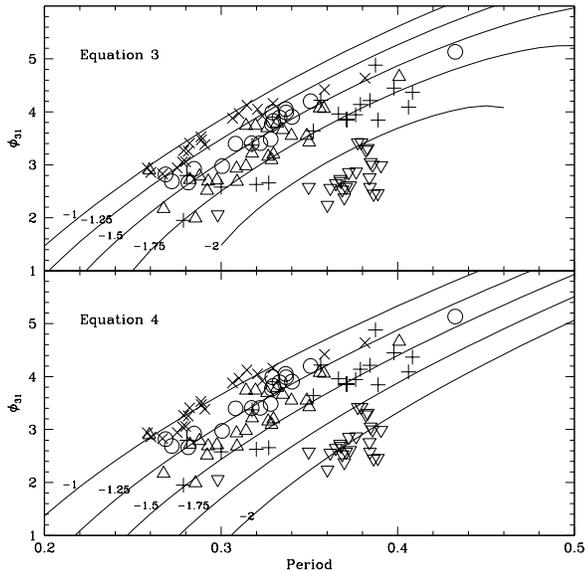}
\caption{Lines of constant [Fe/H], based upon the fits to the ZW84 
(equation~\ref{zwfit}) and the CG97 (equation~\ref{cgfit}) metallicity scales are shown, 
along with the RRc data from Figure~\ref{fig1}.  [Fe/H] values for each line are given
in the lower left corner of each graph.}
\label{fig4}
\end{figure}

\begin{figure}
  \includegraphics[width=84mm]{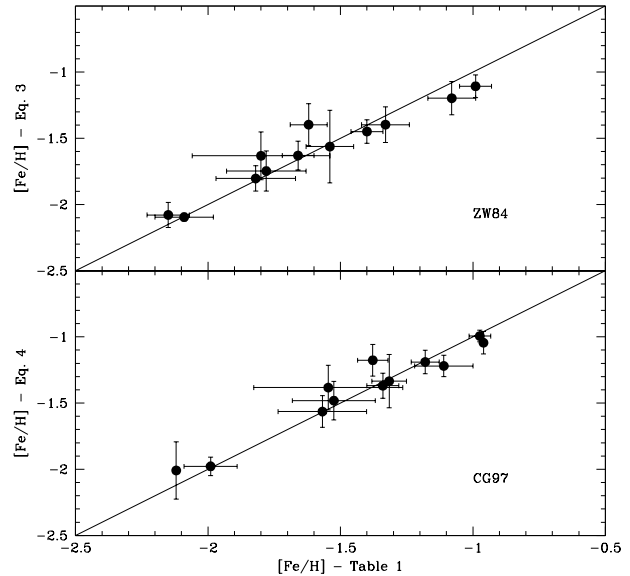}
\caption{Average cluster metallicities based upon ZW84 and equation~\ref{zwfit} (top), and CG97 and 
equation~\ref{cgfit} (bottom) are shown, along with a line of value unity.  
}
\label{fig5}
\end{figure}

Lines of constant value of [Fe/H] based upon equations~\ref{zwfit} and \ref{cgfit} 
are shown in Figure~\ref{fig4}, along with the original data from Figure~\ref{fig1}.
The non-linear nature of equation~\ref{zwfit} limits the range of values where the
function is defined in Figure~\ref{fig4}.  The lines follow the general metallicity 
trends seen in the two metallicity systems.
Average metallicities for all of the variables in each cluster listed in Table~\ref{tbl-1} 
based upon equations~\ref{zwfit} and \ref{cgfit} are shown in Figure~\ref{fig5}. 
Error bars in Figure~\ref{fig5} for
the average calculated metallicities are taken as the standard deviation of the 
[Fe/H] values for the stars in each cluster.  The average cluster 
metallicity based upon equation~\ref{zwfit} for the 12 clusters shown
in Figure~\ref{fig5} varies from the ZW84 values with a mean of $0.08$ dex, while the values based
on equation~\ref{cgfit} vary from the CG97 values with of mean of $0.07$ dex.  
NGC 7089 (M2) is the most displaced cluster in Figure~\ref{fig5}, with two of
its three stars having relatively high calculated metallicities compared to the values of ZW84 and
CG97.  This results in an average metallicity for NGC 7089 that is $0.22$ and $0.20$ dex
greater than the ZW84 and CG97 values respectively.  

\section{Test of the [Fe/H] Relations}

\begin{figure}
  \includegraphics[width=84mm]{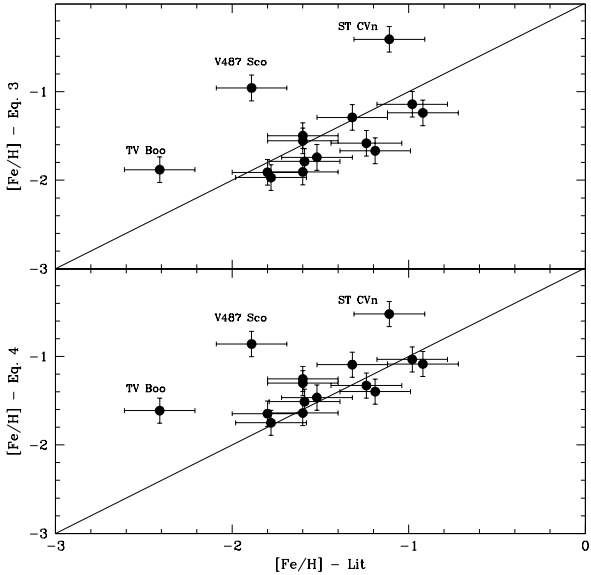}
\caption{Values of [Fe/H] from the literature for field RRc stars compared to values based upon
equations~\ref{zwfit} (top) and \ref{cgfit} (bottom).  TV Boo, V487 Sco, and ST CVn are indicated.  
The line indicates a ratio of unity.}
\label{fig6}
\end{figure}

Equations~\ref{zwfit} and \ref{cgfit} were applied to several test cases to
examine the quality of the relationships in determining [Fe/H] in other
environments.  Field RRc stars were examined first.
The number of field RRc stars with [Fe/H] values and good quality 
$V$ magnitude light curves that could be used is very small, comprised of only 15 stars.  
[Fe/H] values for the stars were taken either from \citet{fb97} or were calculated using 
published $\Delta S$ values and the $\Delta S$ - [Fe/H] relation of \citet{fb97}.  
Some stars have only one measured value of $\Delta S$, while others have 
widely divergent values.  The comparison of the [Fe/H] values from the literature and equations~\ref{zwfit}
and \ref{cgfit} is shown in Figure~\ref{fig6}.  The estimated uncertainty of the individual values of [Fe/H] 
from the literature varies with each source, but typical uncertainties are $0.2$ dex or less.  This
uncertainty is similar to that found in the derivation of equations~\ref{zwfit} and \ref{cgfit} 
(approximately $0.15$ and $0.14$ dex respectively).  These uncertainties are displayed in Figure~\ref{fig6}.
The sample standard deviation of the average metallicity values from
the literature and those derived using our formulae is $0.41$ dex for equation~\ref{zwfit}, 
and $0.42$ dex for equation~\ref{cgfit}.  
There are several stars that are well removed from the
unity relation in Figure~\ref{fig6}.  These are TV Boo, ST CVn and V487 Sco.
TV Boo (${\rm [Fe/H]} = -2.44$) has a metallicity value from the literature outside of the
range used to derive equations~\ref{zwfit} and \ref{cgfit}, which may explain its divergent value.  ST CVn 
($P = 0.329025~{\rm days}$) has a relatively large values of $\phi_{31}$ ($5.11$) for its value of $P$, 
which accounts for its abnormally high calculated values of [Fe/H]. V487 Sco's metallicity  
(${\rm [Fe/H]}=-1.89$) is based upon a single value for $\Delta S$.  When these stars are excluded, 
the sample standard deviations reduce to $0.26$ dex (equation~\ref{zwfit}) and $0.15$ dex (equation~\ref{cgfit}).

\begin{figure}
  \includegraphics[width=84mm]{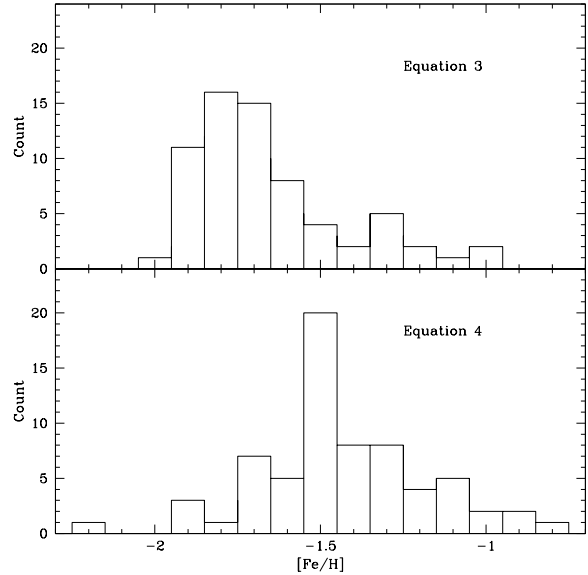}
\caption{The range of metallicity values for RRc stars in $\omega$ Cen, based upon 
equations~\ref{zwfit} (top) and \ref{cgfit} (bottom). }
\label{fig7}
\end{figure}

The RRc stars in $\omega$ Cen were examined next.  Data from the variables in this cluster
were collected from the Fourier coefficient website \citep{m03} and an initial data set was selected
based upon the method of observation, where recent, CCD based coefficients were
favored over photographic observations. These included Fourier coefficients from
\citet{md00}, and \citet{oetal03}.  As with
the globular cluster variables used to derive equations~\ref{zwfit} and \ref{cgfit}, the RRc stars 
in $\omega$ Cen were examined and
poor quality fits or sparse light curves were excluded.  
This reduced the number of variables to $67$, and equations~\ref{zwfit} and \ref{cgfit} were applied to these
stars.  The resulting distribution
of metallicity is shown in Figure~\ref{fig7}.  The relatively wide spread in metallicity
is not surprising given the broad range of metallicity observed in $\omega$ Cen
\citep{soletal05}.  Values that we calculated for [Fe/H] range from $-2.04$ to $-0.97$, with a
mean of $-1.65 \pm 0.24$ using equation~\ref{zwfit}, and a range of $-2.22$ to $-0.83$, with a 
mean of $-1.44 \pm 0.25$ for equation~\ref{cgfit}.  The largest concentration of values is near $-1.77$ and $-1.50$ for
equations~\ref{zwfit} and \ref{cgfit} respectively.  These peaks are near the 
value \citet{soletal05} found for the dominant metallicity population of 
$\omega$ Cen, ${\rm[Fe/H]} \sim -1.7$ dex.

\begin{figure}
  \includegraphics[width=84mm]{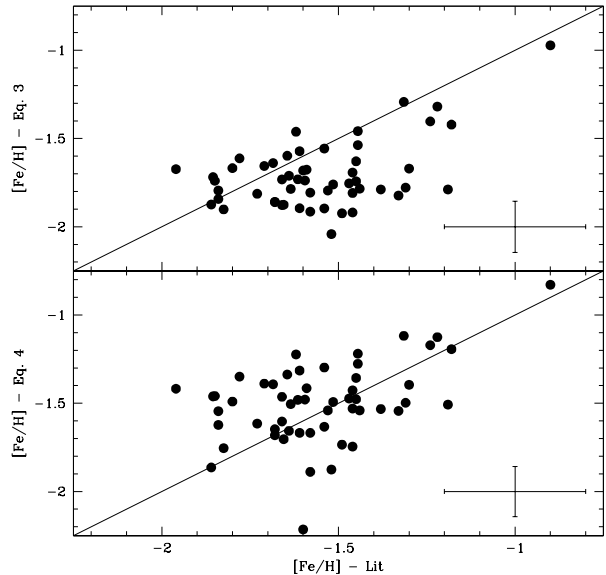}
\caption{Individual RRc stars in $\omega$ Cen with metallicities from the literature, compared to
those based upon equations~\ref{zwfit} (top) and \ref{cgfit} (bottom).  The line indicates a
ratio of unity.  The typical errors associated with the data points are shown.}
\label{fig8}
\end{figure}

There have been several studies that have found the metallicities of individual stars within
$\omega$ Cen using a variety of methods, such as C{\it aby} photometry \citep{rey00}, 
high-resolution spectroscopy \citep{soletal06} and line indices \citep{getal04}.
These methods cover a total of $54$ stars that also have good quality light curves and derived
Fourier coefficients.  The individual stars are compared to the metallicities based on
equations~\ref{zwfit} and \ref{cgfit} in Figure~\ref{fig8}.  The values based upon the ZW84 relation
(equation~\ref{zwfit}) have a sample standard deviation from the published 
metallicity values of $0.25$ dex, while the 
CG97 values (equation~\ref{cgfit}) have a deviation of $0.23$ dex.  Errors for individual stars are
not shown in Figure~\ref{fig8} due to the crowded graph, however most values for errors from the
literature are approximated at $0.2$ dex.  Errors from equations~\ref{zwfit} and \ref{cgfit} are again taken to
be $0.15$ and $0.14$ dex respectively.  

\begin{figure}
  \includegraphics[width=84mm]{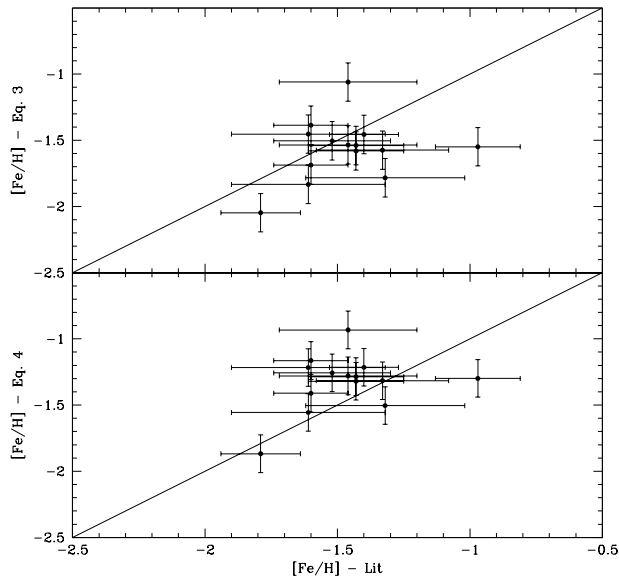}
\caption{Individual RRc stars in the LMC with metallicities taken from \citet{getal04}, compared to
those based upon equations~\ref{zwfit} (top) and \ref{cgfit} (bottom).  The line indicates a
ratio of unity. }
\label{fig9}
\end{figure}

\citet{aetal04} provided $V$-band Fourier coefficients for $682$ RRc in the Large Magellanic
Cloud.  Equations~\ref{zwfit} and \ref{cgfit} were applied to all of these stars and 
respective metallicity averages of ${\rm [Fe/H]} = -1.61 \pm 0.40$ and ${\rm [Fe/H]} = -1.42 \pm 0.37$ were found.  
There were significant deviations from these averages, with some positive metallicity values calculated.  
When stars with large errors ($>0.5$) in their values of $\phi_{31}$ were removed
from the sample, the resulting average metallicity changes slightly, to $-1.66 \pm 0.29$ (equation~\ref{zwfit}) and 
$-1.45 \pm 0.31$ (equation~\ref{cgfit}).  The average [Fe/H] values are very similar to those
of \citet{getal04}, who found a value of $-1.48$ dex for $101$ RR Lyrae, and also
very similar to the metallicity of $23$ LMC RR Lyrae found by \citet{betal04}, ${\rm [Fe/H]}=-1.46$.  
It is possible to do a star-by-star comparison to $14$ 
LMC stars from \citet{getal04}, who calculated
metallicities using line indices.  The statistical comparison to the $14$ stars' metallicities results in sample standard
deviations of approximately $0.28$ for both equation~\ref{zwfit} and \ref{cgfit}, with the 
star-by-star comparison illustrated in Figure~\ref{fig9}.  
It should be noted that the metallicities of individual stars in the LMC and $\omega$ Cen from
the \citet{getal04} study were calibrated using the [Fe/H] scale of \citet{h96}, which is
typically closer in value to the ZW84 system.

The Fourier coefficients for $60$ RRc stars observed in the direction of the galactic bulge 
were derived by MSB98.  These values were obtained from $I$-band photometry, and were 
transformed to the $V$-band using the relations from MSB98.  The average metallicities
obtained using equations~\ref{zwfit} and \ref{cgfit} were $-1.03 \pm 0.50$ and 
$-0.98 \pm 0.37$ respectively.  There are several notable outliers in
the sample though, some of which have been previous noted by MSB98 as likely 
having unusual metallicity values.  These include BW9 V38, and BW11 V55, both of
which have abnormally high values of [Fe/H], and BW1 V11, BW2 V8, BW2 V10, BW4 V46, BW7 V30, BW10 V45 
and BW11 V34, all with at least one value of ${\rm [Fe/H]} < -1.75$.  When these $9$ stars are excluded, 
the metallicity averages become $-0.97 \pm 0.36$ (equation~\ref{zwfit}) and $-0.91 \pm 0.22$ (equation~\ref{cgfit}).
These metallicities compare favorably to the value obtained by \citet{s05} for RRab stars in the
galactic bulge (${\rm [Fe/H]} = -1.04 \pm 0.03$).

\section{Conclusions}

Two formulae were derived that show the relationship between the Fourier coefficient $\phi_{31}$, 
pulsation period and [Fe/H] for RRc stars in $12$ globular clusters.  The formulae are based on the
widely used metallicity scales of \citet{zw84} and \citet{cg97}.  
These relations (equations~\ref{zwfit} and \ref{cgfit})
were able to provide reliable estimates for the value of [Fe/H] for RRc stars in other environments,
including $\omega$ Cen, the LMC, the galactic bulge and field RRc stars.  Even though these results are
encouraging, we do realize that the metallicity relations found here will no doubt evolve as more high
quality light curves for globular cluster RRc stars are made available.  In particular, values of $\phi_{31}$ 
for RRc stars in clusters with very high or very low metallicities would be useful to refine and
improve the formulae derived here, which were based upon cluster [Fe/H] values between approximately 
$-1$ and $-2$.  At the present time it is hoped that the formulae presented here will be of use to 
others investigating the characteristics of RR Lyrae variables.

%

\label{lastpage}

\end{document}